\documentclass[manuscript]{biometrika}

\usepackage{amsmath}
\usepackage{amssymb}
\usepackage{multirow}
\usepackage{times}
\usepackage{bm}
\usepackage{natbib}
\usepackage{color}
\usepackage{pdfpages}

\usepackage[plain,noend]{algorithm2e}

\makeatletter
\renewcommand{\algocf@captiontext}[2]{#1\algocf@typo. \AlCapFnt{}#2} % text of caption
\def\@algocf@capt@plain{top}
\renewcommand{\algocf@makecaption}[2]{%
  \addtolength{\hsize}{\algomargin}%
  \sbox\@tempboxa{\algocf@captiontext{#1}{#2}}%
  \ifdim\wd\@tempboxa >\hsize%     % if caption is longer than a line
    \hskip .5\algomargin%
    \parbox[t]{\hsize}{\algocf@captiontext{#1}{#2}}% then caption is not centered
  \else%
    \global\@minipagefalse%
    \hbox to\hsize{\box\@tempboxa}% else caption is centered
  \fi%
  \addtolength{\hsize}{-\algomargin}%
}
\makeatother

\def\Bka{{\it Biometrika}}
\def\T{^{ \mathrm{\scriptscriptstyle T} }}
\newcommand{\norm}[1]{\left\Vert#1\right\Vert}

\newcommand{\Real}{\mathbb R}

\newcommand{\To}{\longrightarrow}

\def\Var{\mbox{var}}

\def\E{{E}}

\def\To{\longrightarrow}

\newcommand{\Xc}{\mathcal{X}}

\def\1v{ 1}
\def\0v{ 0}
\def\Id{ I}

\def\varphi{\lambda}
\begin{document}

%\jname{Biometrika}
%% The year, volume, and number are determined on publication
%\jyear{xx}
%\jvol{xx}
%\jnum{1}
%% The \doi{...} and \accessdate commands are used by the production team
%\doi{10.1093/biomet/asm023}
%\accessdate{Advance Access publication on 31 July 2012}
\copyrightinfo{\Copyright\ 2012 Biometrika Trust\goodbreak {\em Printed in Great Britain}}

%% These dates are usually set by the production team
%\received{April 2012}
%\revised{September 2012}

%% The left and right page headers are defined here:
\markboth{S. Jung, M. H. Lee \and J. Ahn}{Number of principal components}

%% Here are the title, author names and addresses
\title{On the number of principal components in high dimensions}

\author{Sungkyu Jung}
\affil{Department of Statistics, University of Pittsburgh, Pittsburgh, Pennsylvania, 15260, U.S.A. \email{sungkyu@pitt.edu}}

\author{Myung Hee Lee }
\affil{Center for Global Health, Department of Medicine, Weill Cornell Medicine, New York, New York, 10065, U.S.A.
\email{myl2003@med.cornell.edu}}

\author{\and Jeongyoun Ahn}
\affil{Department of Statistics, University of Georgia,
Athens, Georgia, 30602,
U.S.A. \email{jyahn@uga.edu}}

\maketitle

\begin{abstract}
We consider the problem of how many components to retain in the application of principal component analysis when the dimension is much higher than the number of observations. To estimate the number of components, we propose to sequentially test skewness of the squared lengths of residual scores that are obtained by removing leading principal components. The residual lengths are asymptotically left-skewed if all principal components with diverging variances are removed, and right-skewed if not. The proposed estimator is shown to be consistent, performs well in high-dimensional simulation studies, and provides reasonable estimates in a number of real data examples.
\end{abstract}


\section{Introduction}\label{sec:introduction}
Principal component analysis  is widely used in multivariate analysis,
and has shown to be effective in dimension reduction of modern high-dimensional data.
Let $\Xc = [X_1,\ldots,X_n]\T$ be an $n \times d$ data matrix, where each column vector has zero mean and covariance matrix $\Sigma_d= \sum_{i=1}^d \lambda_i u_i u_i\T$, and $(u_i,\lambda_i)$ denotes the $i$th principal component direction and variance.
 The classical estimates
 $(\hat{u}_i,\hat{\lambda}_i)$
 are obtained by the eigen-decomposition of the sample covariance matrix.
Determining the number of components to retain is a crucial problem in applications of principal component analysis.

A number of strategies have been proposed to tackle this problem in the conventional data situation, where the sample size is large and the dimension is relatively low, {i.e.}, $d \ll n$. These include the graphical methods based on the scree plot of eigenvalues, model-based tests, and computer-intensive tools \citep{Jolliffee2002,josse2012selecting}.
However, modern data challenges often involve the high dimension, low sample size  data situation, i.e., $d \gg n$. Under such situations, those methods may be infeasible, computationally prohibitive, or based on subjective choice. In this article, we propose a novel estimator of the number of components, specifically designed for the $d \gg n$ case.

The true number of components is defined in terms of eigenvalues $\lambda_i$ of  $\Sigma_d$. A  popular approach  is to assume that the first $m$ eigenvalues are larger than a threshold, say $\tau^2$, and the rest of eigenvalues  are equal to $\tau^2$.
This spike model \citep{johnstone2001distribution,Paul2007} has been used in many different contexts \citep{baik2006eigenvalues,kritchman2009non,leek2011asymptotic}.
For diverging dimension $d$ with limited sample size, it has been known that the size of `spike' should be increasing at least at the same order as $d$ in order to have a non-trivial eigenvector estimates \citep{Lee2012a}. Consequently, we assume the eigenvalues of $\Sigma_{d}$ to be
\begin{align}
 \lambda_i &= \sigma_i^2 d \ \ (i = 1,\ldots,m), \quad \sigma_1^2 > \cdots > \sigma_m^2>0, \label{eq:surrogate}
\end{align}
and the rest of eigenvalues $\{\lambda_{m+1},\ldots, \lambda_d \}$ to be equal to each other or form a slowly-decreasing sequence.
%To simplify the discussion, the rest of eigenvalues are assumed to be equal to each other; $\lambda_i = \tau^2$ $(i = m+1,\ldots,d)$. Our general eigenvalue model is much more general, and is defined in Section~\ref{sec:asymptotics}.
%The rest of eigenvalues $\{\lambda_{m+1},\ldots, \lambda_d \}$ are either equal to each other or allowed to be unequal. % as $d\to \infty$, but at a rate much smaller than $d$.
Our model is indeed general, and is defined in Section~\ref{sec:asymptotics}.

%However, the assumption that many eigenvalues are identical to each other $(\lambda_i = \tau^2, i > m$) may appear to be unnatural. Our eigenvalue model is much more general than (\ref{eq:surrogate}), including the case of unequal eigenvalues $\lambda_i > \lambda_{i+1}$, $i>m$. Precise conditions are given in Section~\ref{sec:asymptotics}.

\cite{SJOS:SJOS12264} has shown that under the $m$-component models of (\ref{eq:surrogate}), even though the classical estimates of ($\lambda_i, u_i)$ are inconsistent for increasing dimension, the first $m$ estimated principal component scores contain useful information on the true scores.
 We further show in Section~\ref{sec:scores} that the rest of estimated scores are mostly accumulated noises, which implies that the number of spikes $m$ in (\ref{eq:surrogate}) can be considered as the number of asymptotically meaningful components.

To determine $m$ from a given sample $\Xc$, we propose to sequentially test the null hypothesis
$H_k : m = k$ against the alternative hypothesis $H_{a,k} : m > k$, for increasing values of $k$, and to estimate $m$ by the smallest $k$ for which $H_k$ is not rejected.
To this aim, we show that the squared lengths of
residuals that are obtained by removing the first $k$ leading principal components are asymptotically left-skewed under the null hypothesis, or right-skewed under the alternative hypothesis. This observation motivates us to consider test statistics based on the empirical distribution of the residual lengths. We adopt well-known tests for  skewness \citep{randles1980asymptotically,d1973tests}. The resulting estimator is consistent under a mild condition.

We demonstrate our approach in both simulated and real data sets, including high-dimensional gene expression and image data. In comparison to a number of existing methods, summarized in Section~\ref{sec:Existing methods}, our method provides reasonable estimates. We conclude with a discussion on the effectiveness of estimated  principal components.

\section{Sequential tests to determine $m$}\label{sec:2}

\subsection{Motivation}
We propose to sequentially test the set of null hypotheses $\{H_0,  H_1, \ldots, H_M\}$ for some $M < n$, against one-sided alternatives:
\begin{equation}\label{eq:tests}
H_{k}: m = k  \quad \mbox{  vs.  }\quad  H_{a,k}: m > k,
\end{equation}
where $m$ is the number of components with fast-diverging variances in (\ref{eq:surrogate}).
These null hypotheses do not overlap; if $H_k$ is true, then $H_{\ell}$ is not true for all $\ell \neq k$. However, $H_k$ is nested within all lower-order alternatives; if $H_{k}$ is true, then $H_{a,\ell}$ is true for all $\ell < k$.
These observations suggest to test $H_k$ only if $H_\ell$ is rejected for all $\ell<k$. The number of effective components, $m$, is then determined by the smallest $k$ for which $H_k$ is not rejected at a given level.

%\subsection{Motivation}
To test these hypotheses, we first note that the squared lengths of data vectors $\norm{X_j}_2^2$ $(j =1,\ldots, n)$ have different empirical distributions depending on which, null or alternative, hypothesis is true.
As an illustrative example, let us first assume that the global null $H_0$ is true, specifically, $\Sigma_d = \tau^2 \Id_d$. It can be shown that the squared length $\norm{X_j}_2^2$ is  normally distributed for large $d$:
\begin{equation}\label{eq:normalexample1}
\sqrt{d} \left(d^{-1}\norm{X_j}_2^2 - \tau^2 \right) \sim N(0, 2\tau^4).
\end{equation}
%In fact, we will show in Theorem~\ref{thm:1rateofconvergenceDUALmatrix} that even  without the Gaussian assumption $d^{-1}\norm{X_j}_2^2$ has an asymptotic normal distribution for large dimension $d$.
On the other hand, when $m \ge 1$ in (\ref{eq:surrogate}), it is easy to see that the squared length is decomposed into a sum of two independent random variables:  Assuming $m = 1$ and the data are normal, $d^{-1}\norm{X_j}_2^2 = Z + Y$, where the distributions of $Z$ and $Y$ are approximately %respectively
\begin{equation}\label{eq:normalexample2}
 Z \sim  N(\tau^2, {2\tau^4}/{d}),
 \quad
 {Y}/{\sigma_1^2} \sim \chi^2_1.
\end{equation}
%\begin{equation}\label{eq:normalexample2}
% Z \sim  N(\tau^2, \frac{2\tau^4}{d}),
% \quad
% \frac{Y}{\sigma_1^2} \sim \chi^2_1.
%\end{equation}
In the limit $d \to \infty$, $Z$ degenerates to $\tau^2$,  thus $d^{-1}\norm{X_j}_2^2$ converges in distribution to a shifted-and-scaled chi-square random variable, which is right-skewed.

The above example suggests to consider   test statistics based on the normality or the skewness of the empirical distribution of the squared lengths.
We will show in Section~\ref{sec:asymptotics} that general asymptotic null and alternative distributions of the squared lengths are similar to those in (\ref{eq:normalexample1}) and (\ref{eq:normalexample2}), even under a non-Gaussian assumption.

\subsection{Test statistics}\label{sec:test procedures}

In testing the global null hypothesis, the asymptotic normality, shown in (\ref{eq:normalexample1}) under $H_0$, can be used.
Let ${p}_0^{{N}} = p^{{N}}(\norm{X_1}_2^2, \ldots, \norm{X_n}_2^2)$ be a p-value in testing the normality of  $\norm{X_j}^2_2$.
Intuitively, if a principal component with large variance is present, ${p}_0^{{N}}$ tends to be small, since the empirical distribution becomes right-skewed as shown in  (\ref{eq:normalexample2}).

For testing higher-order hypotheses $H_k$ for $k \ge 1$, we propose to remove the first $k$ estimated principal components from the data.
We use the classical estimates $(\hat{u}_i,\hat{\lambda}_i)$ of the principal component direction vector and variance pair ($u_i, \lambda_i$) ($i = 1,\ldots, n$)
obtained by the eigen-decomposition of the sample covariance matrix $S_d = n^{-1}\Xc\T\Xc = \widehat{U}\widehat\Lambda \widehat{U}\T$.
Denote the scaled squared length of the $k$th residual for the $j$th observation  by
\begin{equation}\label{eq:Rk}
 R_{j}(k) = \frac{1}{d}\norm{X_j - \sum_{i=1}^k \hat{u}_i \hat{u}_i\T  X_j}_2^2,\quad ( j = 1,\ldots,n;\ k = 0,\ldots, M).
 \end{equation}
 The normality test may be adopted in computing p-values for testing $H_k$. %, leading to $ {p}_k^{{N}} = p^{{N}}( R_{1}(k) ,\ldots,  R_{n}(k) )$ ($0 \le k \le M$).
We will show later in Section~\ref{sec:knownPC} that if $\hat{u}_i$ were a consistent estimator of $u_i$ in the $d$-limit for  $i \le k$, then the asymptotic null distribution of ${R}_{j}(k)$  is Gaussian under $H_k$, thus leading to a uniform null distribution of the p-value. % ${p}_k^{{N}}$.

% The normality test may be adopted in computing the p-value for testing $H_k$, leading to $ {p}_k^{\textsc{N}} = p^{\textsc{N}}( R_{1}(k) ,\ldots,  R_{n}(k) )$ ($0 \le k \le M$).
%We will show later in Section~\ref{sec:knownPC} that if $\hat{u}_i$ were a consistent estimator of $u_i$ in the $d$-limit for  $i \le k$, then the asymptotic null distribution of ${R}_{j}(k)$  is Gaussian under $H_k$, thus leading to a uniform null distribution of ${p}_k^{\textsc{N}}$.

There are, however,  limited situations where $\hat{u}_i$ would be consistent for the growing dimension. In fact, under the fast-diverging eigenvalue assumption (\ref{eq:surrogate}) and in the high dimension, low sample size asymptotic scenario, the first $m$ sample principal component directions are \emph{inconsistent}, while the rest are \emph{strongly inconsistent} with their population counterparts \citep{Jung2012,Lee2012a}.
Moreover,
the true principal component variance $\lambda_i$ is often over-estimated by $\hat\lambda_i$ for $i \le m$. Since the sum of squared scores equals the variance, {i.e.}, $n^{-1}\sum_{j=1}^n{ (\hat{u}_i\T X_j)^2} = \hat\lambda_i$, this overestimation affects (\ref{eq:Rk}) in such a way that $R_{j}(k)$ becomes smaller than desired. Thus the empirical distribution of $ R_{j}(k) $ is stochastically smaller than the said asymptotic normal distribution, and becomes left-skewed (or left-tailed). We will revisit this phenomenon in Section~\ref{sec:inconsistent_asymp}.

To incorporate the left-skewed $ R_{j}(k) $, our first choice of
 the test statistic is from a test for skewness. For observations $y_j = {R}_{j}(k)$ ($j =1,\ldots,n$), suppose the distribution of $y_j$ is continuous with unknown median $\theta$.
\cite{randles1980asymptotically} proposed a nonparametric test for symmetry about $\theta$ based on a $U$-statistic with kernel
\begin{equation*}\label{eq:tripleUstat}
f^*(y_i,y_j,y_k) = \mbox{sign}(y_i + y_j - 2y_k) +
                     \mbox{sign}(y_i + y_k - 2y_j) +
                     \mbox{sign}(y_j + y_k - 2y_i).
 \end{equation*}

This test is sometimes called triples test for symmetry. The triples test is an asymptotic test for large $n$, and \cite{randles1980asymptotically} recommended to use its asymptotic normality when $n>20$.
A one-sided test for left- or right-skewed alternatives is also possible \citep[Section 3.9]{hollander2013nonparametric}. For our purpose, the  p-value is obtained by the asymptotic normality for  one-sided triples test with the alternative of  right-skewed distributions, and is denoted by  \begin{equation}\label{eq:triple}
 p_k^{{R}} = p^{{R}}\{{R}_{1}(k),\ldots, {R}_{n}(k)\}.
 \end{equation}

Our second choice of the test statistic is obtained from a test for normality that is specially designed to be sensitive toward skewed alternatives. This test is based on the sample skewness coefficient $b_1  = m_3 / (m_2)^{3/2}$, where $m_r = n^{-1}\sum_{j=1}^n (y_j - \bar{y})^r$. \cite{d1970transformation} suggested a transformation of $b_1$, defining $Z = \delta  \log [ {b_1}/{\lambda} + \{ \left({b_1}/{\lambda} \right)^2 + 1 \}^{1/2} ]$,
where $\delta$ and $\lambda$ are functions of the theoretical moments of $b_1$ in samples from the normal distribution, and in turn are simply functions of the sample size $n$. The distribution of the transformed $Z$, under normal assumptions, is well-approximated by the standard normal, even for a small sample size $n \ge 8$ \citep{d1970transformation,d1973tests}.
%See also \cite{shapiro1968comparative,spiegelhalter1980omnibus}.
Positive $b_1$ and $Z$ indicate  right-skewness, while their negative values indicate left-skewness. For $H_k$ ($k=0,\ldots,M$), the p-value of the skewness test is defined by
\begin{equation}\label{eq:skewness}
p_k^{{D}} = p^{{D}}\{{R}_{1}(k),\ldots, {R}_{n}(k)\} = 1 - \Phi(Z),
 \end{equation}
where $\Phi$ is the distribution function of the standard normal.

Both p-values in (\ref{eq:triple})-(\ref{eq:skewness}) have the uniform distribution if the null distribution of ${R}_{j}(k)$ is normal. The p-values are sensitive to the right-skewed alternatives, as they tend to be close to zero under such cases. On the other hand, if the null distribution of ${R}_{j}(k)$ is left-skewed, the p-values are close to 1.   Other tests of symmetry \citep[\emph{cf.} ][]{farrell2006comprehensive}  can be used in place of (\ref{eq:triple})-(\ref{eq:skewness}).

\subsection{Example}\label{sec:example}

Before proceeding with theoretical results, we demonstrate the proposed test procedures on  a real data set from a microarray study. This data set, described in detail in  \cite{Bhattacharjee2001}, contains
$d = 2530$ genes from $n= 56$ patients in four different lung cancer subtypes.
An inspection of principal component scores plot (see, e.g., Fig.~1 in \cite{Jung2009a}) suggests that the four subtypes are visually separated by using the first few sample principal components, and there is no outlier in the data. It has been believed that the sample principal component analysis provides a reasonable dimension reduction, but there has been no attempt to systematically determine the number of effective components for this dataset.

\begin{figure}
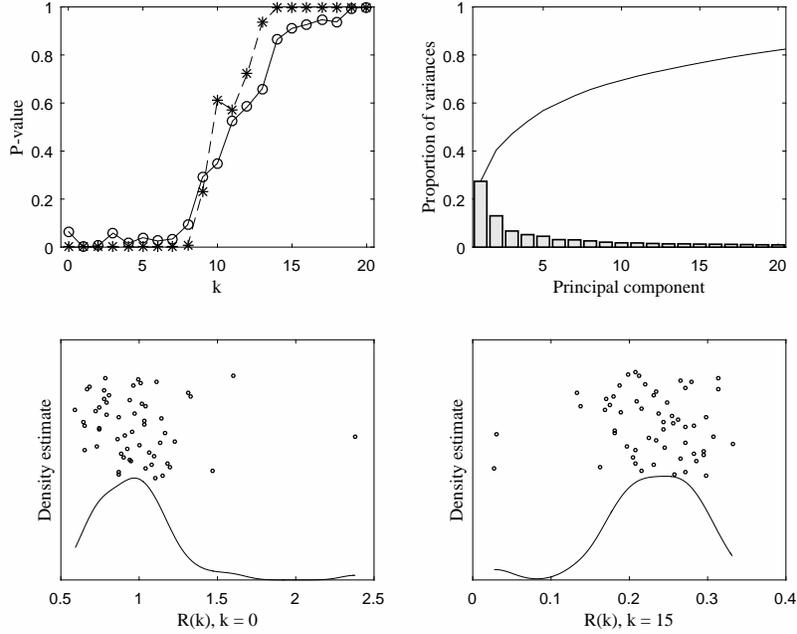

\vspace{0.5em}
\figurebox{20pc}{25pc}{}[Lung1BW.ps]
 \caption{Top left: sequences of p-values computed from a microarray data set \citep{Bhattacharjee2001}; $p_k^{{R}}$ (circle-solid) and $p_k^{{D}}$ (asterisk-dashed). Top right: scree plot. Bottom panels: empirical distributions of $R_{j}(0)$ and  $R_{j}(15)$ by a kernel density estimate and a jitterplot.
 \label{fig:NeilHayesLung3}}
\end{figure}

We applied the tests discussed in Section~\ref{sec:test procedures} to obtain  sequences of p-values in testing (\ref{eq:tests}). As a visual tool to determine the number of components, we plot $p_k^{{R}}$ and $p_k^{{D}}$ against $k$, as shown in the top left panel of Fig.~\ref{fig:NeilHayesLung3}. Graphical methods based on the scree plot, shown in the top right panel of Fig.~\ref{fig:NeilHayesLung3}, lead to either $\hat{m} = 2$ when locating an `elbow,' or $\hat{m} = 17$ when using a heuristic cutoff based on the cumulative percentage of variance explained, say 80$\%$.
In contrast, our estimate, using either of the two test statistics, is $\hat m = 9$, based on
\begin{align}\label{eq:mb_Simple}
\hat{m} = \min \{k : p_k >  \alpha \},
\end{align}
where $\alpha = 0.1$, in this example.

The empirical distribution of $R_{j}(0)$ in the bottom left panel of Fig.~\ref{fig:NeilHayesLung3} is clearly right-skewed, which shows a strong evidence against $H_0: m =0$. The components with large variances contained in $R_{j}(0) = d^{-1} \norm{X_j}^2 $ yield the heavy-tailed distribution skewed to large values.
On the other hand, noise-accumulated components are excessively taken out from $R_{j}(15)$, leading to the left-skewed distribution of residual lengths, as shown in the bottom right panel.
In this example, p-values in the sequences are small for the first few, then show a sharp transition to high values close to 1. This pattern of p-value sequence was also found in many real and simulated data sets, and seems to be typical.

\section{Asymptotic null and alternative distributions}\label{sec:asymptotics}

\subsection{Models}

 We model that the variances of the first $m$ components are fast-diverging at the rate of $d$, while the rest are assumed to be much smaller. It seems standard to assume $m$ is fixed as the variances are diverging.
%The $m$-component model is defined for increasing $d$ in Conditions~\ref{A1}--\ref{A5} below.
For a fixed $m$, the $m$-component model is defined for increasing $d$ in Conditions~\ref{A1}--\ref{A5} below.

\begin{condition}\label{A1}
$ \lambda_i = \sigma_i^2 d$  ($i = 1,\ldots,m$),   $\ \sigma_1^2  > \cdots > \sigma_m^2 >0$.
\end{condition}

\begin{condition}\label{A2-A3}
$\sum_{i=m+1}^d \lambda_i /d \to \tau^2 \in (0, \infty)$, $\sum_{i=m+1}^d \lambda^2_i /d \to \upsilon^2_O \in (0, \infty)$ as $d\to\infty$, and there exists $\delta\in (0,1]$ such that $ \sum_{i=m+1}^d \lambda_i^{2+\delta} = o(d^{1+\delta/2})$.
\end{condition}

%\end{itemize}

%
%\begin{itemize}
%\item[(A1)] $ \lambda_i = \sigma_i^2 d, \ i = 1,\ldots,m, \ \sigma_1^2  > \cdots > \sigma_m^2 >0$.
%\item[(A2)] $\lim_{d\to\infty} \sum_{i=m+1}^d \lambda_i /d = \tau^2 \in (0, \infty)$.
%\item[(A3)] $\lim_{d\to\infty} \sum_{i=m+1}^d \lambda^2_i /d = \upsilon^2_O \in (0, \infty)$, and there exists $\delta\in (0,1]$ such that $ \sum_{i=m+1}^d \lambda_i^{2+\delta} = o(d^{1+\delta/2})$.
%\end{itemize}

By decomposing each observation into the first $m$ principal components and the remaining term, we write $X_j = \sum_{i=1}^m \lambda_i^{1/2} u_i z_{ij} + \sum_{i=m+1}^d \lambda_i^{1/2} u_i z_{ij}$ ($j = 1,\ldots,n$), where $z_{ij}$ is the normalized principal component score.

\begin{condition}\label{A4}
 For each $j = 1,2,\ldots$, ${Z}_{j} = (z_{1j},z_{2j},\ldots)$ is a sequence of independent random variables such that for any $i$, $\E(z_{ij}) = 0$, $\Var(z_{ij}) = 1$, and that the second and third moments of $z_{ij}^2$ are uniformly bounded below and above. For $j \neq \ell$,  ${Z}_{j}$ and ${Z}_{\ell}$  are independent.
\end{condition}

%
%\begin{itemize}
% \item[(A4)]  $\{z_{ij}: i \ge 1, \ 1 \le j \le n \}$ is a sequence of independent random variables such that for all $i$, $\E(z_{ij}) = 0$, $\Var(z_{ij}) = 1$, and that the second and third moments of $z_{ij}^2$ are uniformly bounded below and above.
%\end{itemize}

Conditions~\ref{A1}~and~\ref{A2-A3} are quite general and encompass the spike models of \cite{leek2011asymptotic} and \cite{SJOS:SJOS12264}. In particular, they include
equal ($\lambda_{m+1} = \cdots = \lambda_{d}$),
polynomially-decreasing ($\lambda_i = {i}^{-\beta}$ for $0 \le \beta < 0.5$),
and slowly-diverging eigenvalues ($\lambda_{m+1} = \cdots = \lambda_{m+\nu} = d^\gamma$ and $\lambda_{\nu+1} = \cdots = \lambda_{d} = \tau^2$, where $\nu \asymp d^\beta$, and $2\gamma + \beta < 1$).
%We use the notation $\gamma_k \in [0, 0.5)$ for ${\lambda_k} \asymp {d^{\gamma_k}}$.
%When slowly-diverging $\lambda_k$ is assumed for $k > m$, we denote the rate by $\gamma_k \in [0, 0.5)$, so that ${\lambda_k} \asymp {d^{\gamma_k}}$.
%
%
It is worth noting that Condition~\ref{A2-A3} is stronger than the conditions of \cite{Ahn2007}, which are used in showing the high-dimension, low-sample-size geometric representation: Modulo rotation, the data converge to a regular simplex \citep{Hall2005}.
This stronger condition and the moment conditions in Condition~\ref{A4} are needed in introducing a $d$-asymptotic normality in Theorem~\ref{thm:1rateofconvergenceDUALmatrixNEW} and also in describing asymptotic behaviors of sample scores in Theorem~\ref{prop:samplescore}. We note that Conditions~\ref{A1}~and~\ref{A2-A3} imply a low ``effective rank'' assumption in the random matrix literature; see \cite{koltchinskii2016asymptotics,koltchinskii2017normal} for relevant results.

A special case of our model is the high-dimensional approximate factor model with pervasive factors, defined below, which has recently gained popularity as it is believed to  be more realistic than other models \citep{SJOS:SJOS12264}. Let
$
X = \sum_{i=1}^m q_i z_{i} + \epsilon
$
 be an $m$-factor model, where $z_i$ $(i = 1,\ldots,m)$ are standardized factor scores, $\epsilon =(\epsilon_1,\ldots,\epsilon_d)\T$ is a  mean-zero independent noise vector with uniformly bounded variances.
 The orthogonal factor loadings, $q_i \in \Real^d$, are {pervasive}, that is, the proportion of non-zero entries of $q_i$ is non-vanishing as $d$ increases \citep{fan2013large,SJOS:SJOS12264}. {  For example, $q_i$ is pervasive if for $r \in (0,1)$ the first $\lfloor rd \rfloor$ entries of $q_i$ are one, while the rest $d-\lfloor rd \rfloor$ entries are zero for all $d$.}
 The loading vector $q_i$ is then expressed as $q_i = {\lambda_i^{1/2}} u_i$, where $\norm{u_i}_2 = 1$ and  $\lim_{d\to\infty} \lambda_i /d  = \sigma_i^2$.
% The first $m$ eigenvalue-eigenvector pairs of $\Cov(X)$ are well-approximated by $(\lambda_i, u_i)$, $i\le m$, for large $d$.
%Even if $\Var(\epsilon_i) \neq \Var(\epsilon_j)$ for some pairs $(i,j)$, samples from (\ref{eq:factormodel}) are not distinguishable from samples from (\ref{eq:surrogate}). Thus, the $m$-component model (\ref{eq:surrogate}) can be regarded as the covariance matrix of $X$ .
%
Condition~\ref{A1} makes the first $m$ components  pervasive. Intuitively, when more variables are added into the analysis (i.e., the dimension $d$ increases), these added variables are not simply noises but are correlated with the pervasive factors.

The following assumption on the pervasive factors plays a crucial role in our test procedures proposed in Section~\ref{sec:test procedures}.
\begin{condition}\label{A5}
The third central moment of $z_{ij}^2$ ($i = 1,\ldots,m , j = 1,\ldots, n$) is positive.
\end{condition}
%\begin{itemize}
% \item[(A5)] For the pervasive factors $z_{ij}$ ($i = 1,\ldots,m$), the third central moment of $z_{ij}^2$ is positive.
%\end{itemize}

Simply put, we require that $z^2_{ij}$ is \emph{right-skewed}. We believe Condition~\ref{A5}, which is a relaxation from Gaussian assumption, is quite general. It holds for many known distributions, including $t$-distributions with degrees of freedom $\nu > 6$, the beta distributions with parameters $(\alpha,\alpha)$ with $\alpha > 0.5$, the gamma distributions, and a normal mixture $\xi X_1 + (1-\xi) X_2$, where $X_1$ and $X_2$ are independent normal random variables with a common variance, and $\xi$ follows a Bernoulli distribution.

\subsection{The case of known principal component directions}\label{sec:knownPC}
We  first investigate an ideal case where the principal component directions are known, to better understand the high-dimensional asymptotic behavior of the residual lengths. Define the $k$th \emph{true} residual length of the $j$th observation  by
\begin{align} \label{eq:tildeR}
\tilde{R}_j(k)    = d^{-1}\norm{X_j - \sum_{i=1}^k {u}_i {u}_i\T  X_j}_2^2 = d^{-1} \sum_{i=k+1}^d  w^2_{ij},
\end{align}
where $w_{ij} = u_i\T X_j = \lambda_i^{1/2} z_{ij}$ is the population principal component score.

The asymptotic behaviour of (\ref{eq:tildeR}) can be understood by using a scaled Gram matrix  $S_D = d^{-1}\Xc \Xc\T$, whose $(j,k)$th element is $s_{jk} = d^{-1} X_j\T X_k$, $1\le j,k \le n$. An immediate connection is that the $j$th diagonal element of $S_D$ is $\tilde{R}_{j}(0)$.
Under the assumption of $m$ fast-diverging components, we denote the $n \times m$ matrix of the first $m$ scaled components by
$W_1\T = d^{-1/2} \Xc (u_1,\ldots,u_m)$, where the $(i,j)$th element of $W_1$ is $d^{-1/2} w_{ij}$.
%
%Under the assumption of $m$ fast-diverging components, we denote the $m \times n$ scaled  matrix of the first $m$ components by $W_1 = d^{-1/2} (u_1,\ldots,u_m)\T  \Xc\T$ whose $(i,j)$th element is $d^{-1/2} w_{ij} $.
%
%
It is known that $S_D$ has a limiting expression \citep{Jung2012};
\begin{equation}\label{eq:SdtoS0}
S_D \to  W_1\T W_1 + \tau^2 I_n,\quad d \to \infty,
\end{equation}
in probability, conditional on $W_1$.
This result is now strengthened %in   Theorem~\ref{thm:1rateofconvergenceDUALmatrixNEW}
to provide a rate of convergence of $S_D$.
\begin{theorem}\label{thm:1rateofconvergenceDUALmatrixNEW}
 Assume the $m$-factor model under Conditions~\ref{A1}--\ref{A4}. Let $m \ge 0$ be fixed. Conditioned on the pervasive factors $W_1$, (\ref{eq:SdtoS0}) holds. Moreover,
 each element of $S_D$ has a $d$-asymptotic normal distribution: As $d \to \infty$,
 \begin{align*}
     \sqrt{d}\left( s_{jj} - \sum_{i=1}^m \sigma_i^2 z_{ij}^2 - \tau^2  \right) & \To N(0, \upsilon^2_D) \quad (j = 1,\ldots, n),\\
     \sqrt{d}\left( s_{jk} - \sum_{i=1}^m \sigma_i^2 z_{ij}z_{ik} \right) & \To N(0, \upsilon^2_O) \quad (j,k = 1,\ldots, n; j \neq k)
 \end{align*}
 in distribution, where $\upsilon^2_D = \lim_{d \to \infty }  \sum_{i=m+1}^d \lambda_i^2 \Var(z_{ij}^2) / d$ and $\upsilon^2_O = \lim_{d \to \infty } \sum_{i=m+1}^d \lambda_i^2 /d$.
\end{theorem}

%With the asymptotic normality in Theorem~\ref{thm:1rateofconvergenceDUALmatrixNEW}, we are now ready to specify the null and alternative distributions of $\tilde{R}_j(k)$.
The asymptotic normality in Theorem~\ref{thm:1rateofconvergenceDUALmatrixNEW} provides the null and alternative distributions of $\tilde{R}_j(k)$.

\begin{corollary}\label{cor:result1} Assume the $m$-factor model  under Conditions~\ref{A1}--\ref{A5}. Let $n > m \ge 0$ be fixed. Then for any $j = 1,\ldots,n$ and $k=0,1,\ldots,n-1$, for large $d$,
\begin{itemize}
\item[(i)] if $k \ge m$, $\tilde{R}_j(k)$ is asymptotically normal.
\item[(ii)] if $k < m$, $\tilde{R}_j(k)$ is asymptotically right-skewed.
\end{itemize}
\end{corollary}

Intuitively, if all of the pervasive factors are removed in the residual, {i.e.}, $k \ge m$, then the factors in the residual can be thought of as an accumulated noise, and by Theorem~\ref{thm:1rateofconvergenceDUALmatrixNEW}, the residual length has a limiting normal distribution. On the other hand, if one or more pervasive factors remain in the residual, that is,  $k <m$, then the sum of squared factors appears in the residual length. Condition~\ref{A5} ensures that the squared factors are right-skewed.

\subsection{The case of estimated principal component  directions} \label{sec:inconsistent_asymp}

When the estimated principal component  directions $\hat{u}_i$ are used,  the residual lengths ${R}_j(k)$  have different limiting distributions than those of $\tilde{R}_j(k)$.
We characterize the full family of asymptotic distributions of ${R}_j(k)$ under the null and alternative hypotheses.
For this, we consider an asymptotic scenario where the limits $d\to\infty$ and $n\to\infty$ are taken progressively. This resembles the case where the dimension increases at a much faster rate than the sample size does, such as $d/n \to \infty$, but is not identical to it \citep{lee2014convergence}.
Asymptotic null distributions of $R_j(k)$ for fixed $n$ are discussed in the supplementary material.
%In the supplementary material, as a supplementary theory, we also show  asymptotic null distributions for fixed $n$.

%For this, we consider an asymptotic scenario where both the dimension $d$ and the sample size $n$ diverge. Since the $d$-asymptotic argument is a driving factor in the analysis, we require that the dimension increases at a much faster rate than the sample size does. In particular we assume that $d/n \to \infty$.

Let $\hat{w}_{ij} = \hat{u}_i\T X_j$ denote the sample projection score.
The following decomposition is useful in explaining the limiting distribution of ${R}_j(k)$:
\begin{align}
{R}_j(k)  =  \tilde{R}_j(k) + a_j(k), \quad a_j(k) =  \frac{1}{d}\sum_{i=1}^k (w_{ij}^2 - \hat{w}_{ij}^2). \label{eq:R_decomp}
\end{align}

First consider the asymptotic null distribution of ${R}_j(k)$ under $H_k: m = k$.
The over-estimated $\lambda_i$ ($i =1,\ldots, m$) by $\hat\lambda_i$ leads that $\hat{w}_{ij}^2$ tends to be larger than $w_{ij}^2$ (shown in the supplementary material). Thus one can expect that $a_j(m)$, the difference between the squared true score and sample score, is negative. It turns out that $a_j(m)$, and subsequently ${R}_j(m)$, are in fact left-skewed in the limit, as shown in the following theorem.
Describing the alternative distribution of $R_j(k)$ for $k < m$ seems a bit more challenging, at first glance.  This is because the two dependent variables in (\ref{eq:R_decomp}) exhibit different skewness; the first term, $\tilde{R}_j(k)$, is asymptotically right-skewed, and the second term, $a_j(k)$, is asymptotically left-skewed. Below, we also show that $a_j(k)$ is in fact asymptotically negligible.
%This is because ${R}_j(k)$, $k < m$, is decomposed by (\ref{eq:R_decomp}) into a sum of two dependent variables, where the first term, $\tilde{R}_j(k)$, is asymptotically right-skewed, and the second term, $a_j(k)$, is asymptotically left-skewed. We show that when both $d,n$ diverge, $a_j(k)$ is in fact asymptotically negligible.

\begin{theorem}\label{thm:result02}
Assume the $m$-factor model  under Conditions~\ref{A1}--\ref{A5}. Let $m \ge 0$ be fixed.
Suppose that the limits $d \to \infty$ and $n \to \infty$ are taken progressively.

\begin{itemize}
\item[(i)] (Global null) If $m = 0$, then in the limit, for each $j$, ${R}_j(0)$ is normally distributed, and for $j \neq \ell$, ${R}_j(0)$ and ${R}_{\ell}(0)$ are independent.
\item[(ii)] (Non-trivial null)
If $m \ge 1$, then for each $j$, $n (R_j(m) - \tau^2) \to A_j(m)$ in probability, where $A_j(m) = - \tau^2 \sum_{i=1}^m z_{ij}^2$. Moreover, the random variables $A_j(m)$ $(j = 1,2,\ldots)$ are identically distributed, left-skewed and independent with each other.
\item[(iii)] (Alternative) If $m>k\ge 0 $, then for each $j$,
  $ R_j(k) \to B_j(k,m)$ in probability, where $B_j(k,m) = \sum_{i=k+1}^m \sigma_{i}^2z_{ij}^2 + \tau^2$. Moreover,
   the random variables $B_j(k,m)$ $(j = 1,2,\ldots)$ are identically distributed, right-skewed and independent with each other.
\end{itemize}
%
%\begin{itemize}
%\item[(i)] (Global null) If $m = 0$, then in the limit, for each $j$, ${R}_j(0)$ is normally distributed, and for $\jmath \neq j$, ${R}_j(0)$ and ${R}_{\jmath}(0)$ are independent.
%\item[(ii)] (Non-trivial null)
%If $m \ge 1$, then for each $j$, $n (R_j(m) - \tau^2) = A_j(m)+ O_p(n^{-1/2}) + O_p(n/d)$, where $A_j(m) = - \tau^2 \sum_{i=1}^m z_{ij}^2$. Moreover, the limiting random variables $\{A_j(m) : j = 1,2,\ldots \}$ are identically distributed, left-skewed and independent with each other.
%\item[(iii)] (Alternative) If $m>k\ge 0 $, then for each $j$,
%  $ R_j(k) = B_j(k,m) + O_p(d^{-1/4}) + O_p(n^{-1/2})$, where $B_j(k,m) = \sum_{i=k+1}^m \sigma_{i}^2z_{ij}^2 + \tau^2$. Moreover,
%   the limiting random variables $\{B_j(k,m) : j = 1,2,\ldots \}$ are identically distributed, right-skewed and independent with each other.
%\end{itemize}
\end{theorem}

Theorem  \ref{thm:result02} provides a theoretical justification for the test procedures  based on the skewness in Section~\ref{sec:test procedures}. The test statistics in (\ref{eq:triple})-(\ref{eq:skewness}) tend to be large under the non-trivial null hypothesis, and small under the alternative.
Theorem  \ref{thm:result02} also implies that the sharp transition of p-values, from low to high, as shown in Fig.~\ref{fig:NeilHayesLung3} is bound to happen for large enough $d$ and $n$.

 Our next result shows that our estimator (\ref{eq:mb_Simple}) consistently estimates the true number of principal components. For this, we require the test involved be consistent and the function $p_k$ be continuous for  each $n$. These hold if $p_k^{{D}}$ is used. Although $p_k^{{R}}$ does not satisfy the continuity condition, the estimator of $m$ using the triples test appears to be consistent in our empirical results.

 %If $p_k^{{D}}$ is computed from  left-skewed (or right-skewed) random variables, then $p_k^{{D}} \to 1$ (or 0, respectively) in probability as the sample size increases. That is, the test of symmetry involved is consistent.

%For this, we require the test of symmetry involved is consistent, which is true when $p_k^{{D}}$ is used; if $p_k^{{D}}$ is computed from  left-skewed (or right-skewed) random variables, then $p_k^{{D}} \to 1$ (or 0, respectively) in probability as the sample size increases. For $p_k^{{R}}$, the hypothesized parameter is the expected value $\eta$ of the U-statistic (\ref{eq:tripleUstat}). Some asymmetric distributions may correspond to $\eta = 0$ (thus no power for such distributions), but as argued in \cite{randles1980asymptotically}, ``the class of asymmetric distribution for which $\eta = 0$ is small.''

%Our next result shows that our estimator (\ref{eq:mb_Simple}) consistently estimates the true number of principal components. For this, we require the test of symmetry involved is consistent, which is true when $p_k^{{D}}$ is used; if $p_k^{{D}}$ is computed from  left-skewed (or right-skewed) random variables, then $p_k^{{D}} \to 1$ (or 0, respectively) in probability as the sample size increases. For $p_k^{{R}}$, the hypothesized parameter is the expected value $\eta$ of the U-statistic (\ref{eq:tripleUstat}). Some asymmetric distributions may correspond to $\eta = 0$ (thus no power for such distributions), but as argued in \cite{randles1980asymptotically}, ``the class of asymmetric distribution for which $\eta = 0$ is small.''

\begin{theorem}\label{thm:result03}
Assume the $m$-factor model  under Conditions~\ref{A1}--\ref{A5}. %Let $m > 0$ be fixed. Suppose that both $d,n \to \infty$ with $d/n \to \infty$.
    Let $\hat{m}(\alpha)$ be the estimator of $m$ defined in (\ref{eq:mb_Simple}), where $p_k$ is computed using (\ref{eq:skewness}). Then for any $\alpha \in (0,1)$,
    $$\lim_{n\to\infty}
    \lim_{d\to\infty} { pr}( \hat{m}(\alpha) = m ) = 1.$$
    \end{theorem}

Theorem~\ref{thm:result03} not only shows a consistency but also suggests that for large enough dimension and sample size, the estimator $\hat{m}$ should be nearly invariant to different choices of $\alpha$. This robustness against varying $\alpha$ is empirically confirmed in Section~\ref{sec:choice_of_alpha}.

\section{Numerical studies}\label{sec:Numerical studies}

\subsection{Existing methods to compare}\label{sec:Existing methods}
There are a number of existing methods for determining the number of components. For $d \ll n$ case, we refer to \cite{Jolliffee2002} for an extensive list and discussion of heuristic and model-based methods. % that appear to work well in the  $d \ll n$  case.
%Computer-intensive methods have also been proposed,
%and these include cross-validation and permutation  methods  \citep[\emph{cf}. ][and references therein]{Jolliffee2002,diana2002cross,josse2012selecting} and a permutation-based method of \cite{buja1992remarks}.
%and these include cross-validation methods  \citep[\emph{cf}. ][and references therein]{Jolliffee2002,diana2002cross,josse2012selecting} and a permutation-based method of \cite{buja1992remarks}.

\cite{bai2002determining} considered the problem of determining the number of principal components, $m$, when both the dimension and sample size diverge. There are several information-criteria type estimators proposed in their work, but we found directly using these estimators yield unsatisfactory results in our experiments. For our empirical studies, we use a modified estimator based on their information criteria, defined in the supplementary material.
Simulation-based methods such as parallel analysis \citep{horn1965rationale} have evolved into eigenvalue-based estimations of $m$, using an asymptotic random matrix theory for large $d$ and $n$.   \cite{kritchman2008determining,kritchman2009non} and  \cite{passemier2012determining,passemier2014estimation} %and \cite{patterson2006population}
developed estimators of $m$ using Tracy-Widom distribution \citep[\emph{cf.} ][]{johnstone2001distribution}.
\cite{leek2011asymptotic} also proposed an eigenvalue-based estimator of $m$ by choosing a stable threshold for the sample eigenvalues.

Our estimators obtained by (\ref{eq:mb_Simple}) will be denoted by $\hat{m}_R$ and $\hat{m}_D$, when using the p-value sequences of (\ref{eq:triple}) and (\ref{eq:skewness}), respectively. For simplicity, we have used $\alpha = 0.1$ for all numerical results. Our methods are in fact robust to different choices of $\alpha$, as further discussed in Section~\ref{sec:choice_of_alpha}. In the numerical studies below, the performances of our estimators are compared with the methods of
\cite{kritchman2008determining}, \cite{passemier2014estimation}, \cite{leek2011asymptotic} and \cite{bai2002determining}.
%$\hat{m}_{KN}$, $\hat{m}_{PY}$, $\hat{m}_L$ and $\hat{m}_{BN}$.

\subsection{Real data examples}\label{sec:realdata}
We report the estimated number of components for eight real data sets. The first six are from  gene expression studies, which usually produce high-dimensional data with limited sample size. The latter two data are two different types of images. These data sets are described in the following, and the result is summarized in Table~\ref{tab:realdata}.

\begin{table}
\def~{\hphantom{0}}
\tbl{Number of principal components estimated from real data sets}{%
%\tbl{Number of principal components estimated from real data sets. The names of data sets are followed by dimension and sample size $(d,n)$ of the data, and estimates $\hat{m}_{AD}$ , $\hat{m}_{T}$   and $\hat{m}_D$  from (\ref{eq:mb_Simple}) , $\hat{m}_L$ of \cite{leek2011asymptotic}, $\hat{m}_{KN}$  of \cite{kritchman2008determining}, $\hat{m}_{PY}$ of \cite{passemier2014estimation}, and $\hat{m}_{BN}$ for our modified estimator based on \cite{bai2002determining}. See text for explanation of data used in the analysis.}{%
 \begin{tabular}{cccccccccc}
   Type & Data set & $(d,n)$ &  $\hat{m}_{R}$  & $\hat{m}_D$  & $\hat{m}_L$ & $\hat{m}_{KN}$ & $\hat{m}_{PY}$ & $\hat{m}_{BN}$ \\
%\hline
\multirow{6}{*}{Gene} &   DLBCL  & (7129, 77) &11&11&31& 65 & 65 & 10   \\
   & Prostate  & (2135, 102)  & 22 &22 &14 & 52 & 25 & 14 \\
   & NCI60 (cDNA)  & (2267, 60) & 5& 5& 2& 31 & 9 & 2 \\
   & NCI60 (Affy)  & (2267, 60) & 10& 10&23 &44 & 31 & 4 \\
   & NCI60 (combined)  & (2267, 120) &11 &11 &65 & 86 & 80 & 4  \\
   & Leukemia        & (3051, 38)   &1 &9  &9 & 25 & 22 & 3\\
   & Lung            & (2530, 56)   & 9& 9& 55& 41 & 31 & 7\\
   & Lobular Freeze  & (16615,817)  & 118 & 92 & 20 & 481 & 171 & 29 \\
%\hline
\multirow{2}{*}{Image}   & Hippocampi & (336, 51)  & 11 & 11& 14& 27 & 24 & 3  \\
                         & Liver      & (12813, 500) & 71 & 151& 171& 416 & 290 & 137\\
  \end{tabular}}
\label{tab:realdata}
\begin{tabnote}
$\hat{m}_R$ our estimator using (\ref{eq:triple}); $\hat{m}_D$ our estimator using (\ref{eq:skewness}); $\hat{m}_L$, \cite{leek2011asymptotic} method; $\hat{m}_{KN}$, \cite{kritchman2008determining} method; $\hat{m}_{PY}$, \cite{passemier2014estimation} method; $\hat{m}_{BN}$, \cite{bai2002determining} method.
\end{tabnote}
\end{table}

The microarray data sets we tested include Diffuse large B-cell lymphoma data \citep[\emph{DLBCL}, ][]{shipp2002diffuse}, \emph{Prostate} cancer  data \citep{singh2002gene}, and each of the two different platforms of the \emph{NCI60} cell line data \citep{shoemaker2006nci60}. % (available at \verb"http://discover.nci.nih.gov/" and see also \cite{shen2008sparse}). The number of components is estimated for each platform separately, and also for combined data.
We also tested the training set of \emph{Leukemia} data \citep{Golub1999} and \emph{Lung} cancer data \citep{Bhattacharjee2001}. The \emph{Lobular Freeze} data set is a breast cancer gene expression data, measured by RNA sequencing \citep{ciriello2015comprehensive}.

The \emph{hippocampi} data set \citep{Pizer2011} consists of skeletal-representations, 3-dimensional models of human organs, parameterized by spatial locations, lengths and directions of skeletal spokes.  \cite{Pizer2011} proposed a non-classical principal component analysis based on \cite{JungPNS} to handle the complex data set. This data example is chosen to show that our method can be applied to a non-classical principal component analysis through   scores matrix, since the residual lengths can be computed from the scores; see (\ref{eq:R_decomp}). %Note that the eigenvalue-based estimator $\hat{m}_{KN}$ is well-defined here, while this comparison is not fair to $\hat{m}_{L}$, which needs the full data set to determine the stable threshold.
The last data set consists of cell nucleus grayscale images from human \emph{liver} tissues \citep{Wang2011b}. we chose $d= 12813$ variables with standard deviation greater than 0.01 from  the original $36864$ pixels.
%For this data set, the sample size is relatively large ($n = 500$), and the estimated numbers of effective PCs are also in general large: $\hat{m}_R = 71$ and $\hat{m}_D = 151$. For large sample sizes, the number of effective PCs tends to be large. As we decrease the sample size (by choosing a random subset of the sample) the estimates of $m$ decreases for all five methods we tested.

The estimates in Table~\ref{tab:realdata} show that our estimates $\hat{m}_R$ and $\hat{m}_D$ are similar to each other. There is a clear tendency that our estimates are generally larger than the estimates of \cite{bai2002determining}, but smaller than the estimates of \cite{kritchman2008determining} and \cite{passemier2014estimation}.
Through a simulation study (summarized in Section~\ref{sec:sim_results}), we have come to believe that the method of Bai \& Ng tends to underestimate, while Kritchman \& Nadler and Passemier \& Yao overestimate, for finite $d$ and $n$.
In particular, the estimates $\hat{m} = 25$ and $ 22$ from the methods of Kritchman \& Nadler and Passemier \& Yao for the leukemia data seem unsuitably large considering the sample size $n = 38$.
Our estimates, especially $\hat{m}_D$, exhibit a balance between two extremes.
The seemingly biased other estimates are in part caused by the violation of distributional assumptions such as normality and equal tail-eigenvalues, which might be the case for real data sets. Our estimates do not need such assumptions.
%As an anonymous referee pointed out, the empirical eigenvalues are sufficient for the estimation of $m$ only under the normal distributional assumption. While $\hat{m}_{KN}$ and $\hat{m}_{PY}$ and $\hat{m}_{L}$ only use the empirical eigenvalues, our estimates and $\hat{m}_{BN}$ make use of the residual.
%Finally, our estimates are robust to different choices of $\alpha$ as exemplified in Section~\ref{sec:choice_of_alpha}.

\subsection{Simulation}\label{sec:sim_results}
%Although $\hat{m}_{KN}$ and $\hat{m}_{PY}$ consistently estimate $m$ under the equal-eigenvalue assumption, they tend to overestimate under our general assumptions. Moreover, the quality of the principal subspace estimation for the dimension $\hat{m}_{KN}$ will be poor. For example, in the Leukemia data case, it is almost impossible with the limited sample size $n = 38$ to correctly estimate the subspace of dimension $\hat{m}_{KN}= 25$.

To better understand the empirical performances of the estimators, we conducted a simulation study. The eigenvalues of $\Sigma_d$ are modeled using $s>0$ representing a signal strength, $0 \le \beta < 1/2$ representing a decay rate of variances in noise components, and $g>0$ controlling the gap between leading eigenvalues, by
%
%For $s>0$ representing a signal strength, $0 \le \beta < 1/2$ representing a decay rate of variances in noise components, and $g>0$ controlling the gap between leading eigenvalues, we set the eigenvalues of $\Sigma_d$ as
\begin{equation}\label{eq:Sigmamodel}
\lambda_i = \left\{
              \begin{array}{ll}
                \sigma^2_i d,\quad \sigma_i^2 = s^2\{1+ g(m-i)\}, & {i = 1,\ldots, m ;} \\
                \tau_\beta i^{-\beta},                        & {m < i \le d,}
              \end{array}
            \right.
\end{equation}
where $\tau_\beta = \{\sum_{i=m+1}^d i^{-\beta}/ (d-m)\}^{-1}$ is used to ensure that the average of $\lambda_i$ ($i=m+1,\ldots,d$) is 1. The eigenvectors of $\Sigma_d$ are randomly chosen from the uniform distribution on the orthogonal group of dimension $d$.
%All models satisfy (A1)--(A6).
%We experimented three models, standard normal,  uniform on $(-\sqrt{3},\sqrt{3})$ and a scaled $t_{7}$ for sampling of the standardized scores $z_{ij}$.

We present simulation results for four different cases:
\begin{itemize}
\item Case I: The standard normal distribution is used for sampling of the standardized scores $z_{ij}$. The eigenvalues of population covariance matrix is defined by (\ref{eq:Sigmamodel}) with $(s,g,\beta) = (0.2, 1, 0)$.
\item Case II: The standard normal model with $(s,g,\beta) = (0.2, 1, 0.3)$.
\item Case III: The $t_3$ distribution model with $(s,g,\beta) = (0.2, 1, 0.3)$.
\item Case IV: The $t_3$ distribution model with $(s,g,\beta) = (0.1, 0.5, 0.3)$.
\end{itemize}

\begin{figure}
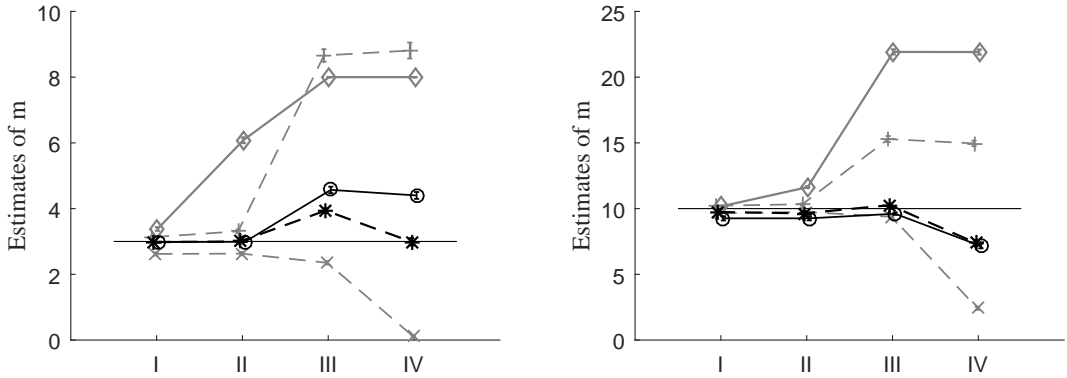

\figurebox{12pc}{}{}[SIM_collection_ppC1.ps]
   \caption{Left: $m = 3$ model. Averages and standard errors of estimates of the number of components; $\hat{m}_R$ (circle-solid), $\hat{m}_D$ (asterisk-dashed), \cite{kritchman2008determining} estimator (gray plus-dashed), \cite{passemier2014estimation} estimator (gray diamond-sold) and \cite{bai2002determining} estimator (gray x-dashed). The estimates of \cite{leek2011asymptotic} were similar to \cite{bai2002determining}, thus omitted. Right: $m = 10$ model. The largest standard error for all results is 0.25.
    \label{fig:SIM_collection_C1}}
\end{figure}

We set the true number of components $m= 3, 10$ for each of the cases and collected the estimated results for $(d,n) = (10000,100)$ based on 100 simulation runs.  The results in Fig.~\ref{fig:SIM_collection_C1} show that our estimators $\hat{m}_D$ and $\hat{m}_R$ perform as well as or better than the competing estimators.

Case I is an ideal situation for all methods considered. In particular, the variances of noise components are equal to each other ($\lambda_{i} = 1$, for all $i = m+1,\ldots,d$), and the normal assumption is satisfied. All methods perform similarly.
In the settings with slowly-decreasing tail-eigenvalues (Case II), the methods of \cite{kritchman2008determining} and \cite{passemier2014estimation} tend to overestimate. This is because, for $\beta>0$, the equal tail-eigenvalue assumption for the estimators  of Kritchman \& Nadler and Passemier \& Yao to be consistent is not satisfied. %The methods of \cite{bai2002determining} and \cite{leek2011asymptotic} often lead to underestimation of $m$.
The assumptions for consistency of our estimators are satisfied under Case II.

In cases III and IV, a scaled $t$ distribution with degrees of freedom $3$ is used to sample the standardized scores $z_{ij}$. The coefficient of skewness  in Condition~\ref{A5} is not defined for this heavy-tailed distribution. Nevertheless, our estimators are less affected by the violation of the assumption compared to the more biased estimators of Kritchman \& Nadler and Passemier \& Yao. This is because the heavy-tailed scores exhibit more drastic distinctions of the left- and right-skewness than using the normal distribution. In Case IV, the leading $m$ eigenvalues are smaller than in Case III. In this weak signal setting,  both estimators of \cite{leek2011asymptotic} and \cite{bai2002determining} are severely underestimating.

More simulation results, including larger $d$, larger $m$, various settings of $(s,g,\beta)$, and the conventional $n>d$ case, are reported in the supplementary material. %In particular, we report that our methods, $\hat{m}_D$ and $\hat{m}_R$, are shown to perform better for larger dimension $d$, which is expected from the theoretical results, and that  our methods perform better for larger signal $s$, and are robust against different values of $\beta$.

\subsection{Empirical robustness against varying $\alpha$}\label{sec:choice_of_alpha}

The asymptotic invariance of $\hat{m}$ against varying $\alpha \in (0,1)$, shown in Theorem~\ref{thm:result03}, suggests some degrees of invariance for moderately large $d$ and $n$. In fact, for most real and simulated data examples we considered, the values of $\hat{m}$ are stable against various values of threshold $\alpha$. %Two examples are shown here.

%Although $\hat{m}$ is asymptotically invariant for varying $\alpha \in (0,1)$ as shown in Theorem~\ref{thm:result03}, it may not be the case in practice where the sample size and the dimension are finite. We, however, have observed that for most data examples considered for modestly large $d$ and $n$ the values of $\hat{m}$ are stable against various values of ``threshold'' $\alpha$. Two examples are shown here.
%in practice where the sample size and the dimension are finite, it may be the case that different choices of $\alpha$ may produce different estimates $\hat{m}$.
%In this section, we show an examples that for modestly large $d$ and $n$ the values of $\hat{m}$ are stable against various values of ``threshold'' $\alpha$.

\begin{figure}
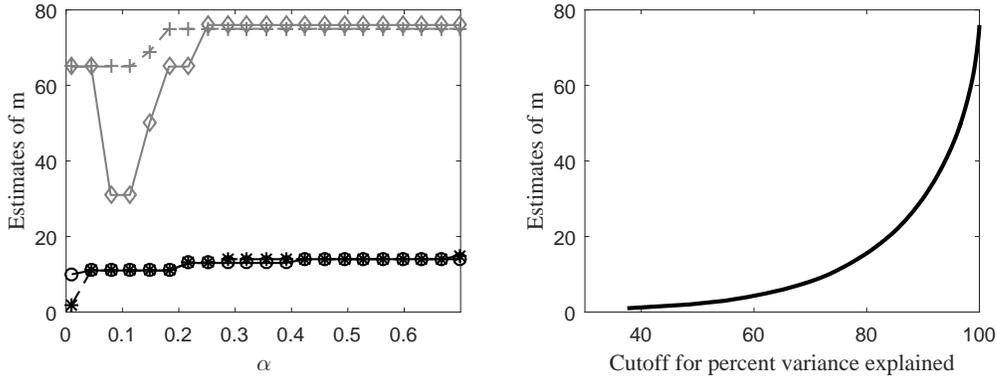

\vspace{1em}
% The arguments in the next line are {height}{optional width}{used only by OUP for typesetting}[filename, in directory art]
\figurebox{12pc}{}{}[Figure3RobustToAlpha_pDLBCL.ps]
% note that files may not be rotated
   \caption{Estimates $\hat{m}$ as functions of $\alpha$ (left panel) or as a function of the variance threshold (right panel), computed from a real data set of \cite{shipp2002diffuse};  $\hat{m}_R$ (circle-solid), $\hat{m}_D$ (asterisk-dashed), \cite{kritchman2008determining} estimator (gray plus-dashed), \cite{passemier2014estimation} estimator (gray diamond-sold). \label{fig:Figure3RobustToAlpha_DLBCL}}
\end{figure}

For a real data set of \cite{shipp2002diffuse}, introduced in Section~\ref{sec:realdata}, it is confirmed  that our estimates $\hat{m}_R(\alpha)$ and $\hat{m}_D(\alpha)$ are stable for a wide range of $\alpha$; see Fig.~\ref{fig:Figure3RobustToAlpha_DLBCL}.
As a comparison, we also have experimented on the eigenvalue-based estimators of \cite{kritchman2008determining} and \cite{passemier2014estimation} by changing their threshold value, which is parameterized by the 1-$\alpha$ quantiles of Tracy-Widom distribution. These estimates change their values  more substantially.
(The methods of \cite{leek2011asymptotic} and \cite{bai2002determining} are not subject to arbitrary choices of threshold, thus  excluded from this study.) We further compare to a heuristic method using the cumulative percentage of variance explained. As shown in the right panel of Fig.~\ref{fig:Figure3RobustToAlpha_DLBCL}, changing the threshold, say from 80\% to 90\%, would drastically change the estimates.

\begin{figure}
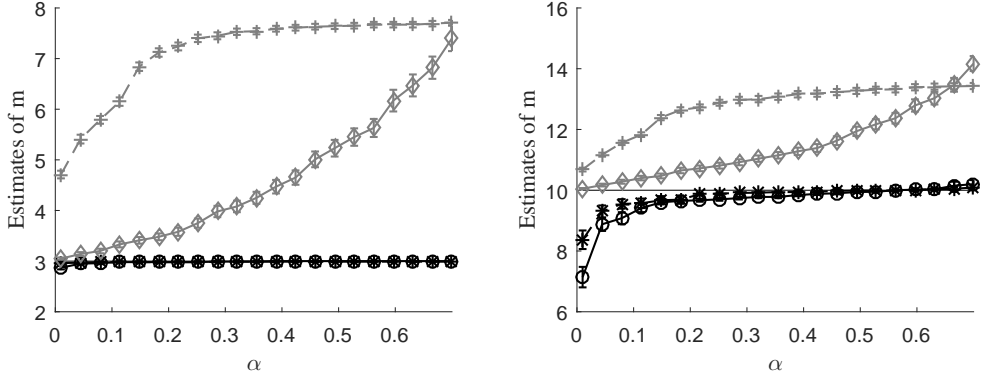

\vspace{1em}
% The arguments in the next line are {height}{optional width}{used only by OUP for typesetting}[filename, in directory art]
\figurebox{12pc}{}{}[Figure3RobustToAlpha_pSIM.ps]
% note that files may not be rotated
   \caption{Estimates as functions of $\alpha$, compared with the true values of $m = 3$ (left) and $10$ (right). Shown are the averages with standard errors from 100 simulation runs; $\hat{m}_R$ (circle-solid), $\hat{m}_D$ (asterisk-dashed), \cite{kritchman2008determining} estimator (gray plus-dashed), \cite{passemier2014estimation} estimator (gray diamond-sold) \label{fig:Figure3RobustToAlpha_SIM}}
\end{figure}

The robustness of our estimators against varying $\alpha$ is also confirmed in simulated data. In
Fig.~\ref{fig:Figure3RobustToAlpha_SIM}, the estimates with varying $\alpha$ are plotted for data generated by Case II in Section~\ref{sec:sim_results}. Our estimates are stable, except for small values of $\alpha<0.1$. The estimator of Kritchman \& Nadler is also stable for  larger values of $\alpha$, but the corresponding estimates are clearly biased.

%\section*{Acknowledgements}
%The authors thank the editor, the associate editor and two referees for constructive comments, and Professor J. S. Marron for sharing the Lobular Freeze data.

\section{Principal component scores in high dimension} \label{sec:scores}
We conclude with a formal statement on the usefulness of the sample principal component scores in high dimensions.

% We show below that under the $m$-component model, the first $m$ sample principal scores are `faithful'  to the corresponding population scores in the limit $d\to\infty$, even though the principal component estimates $(\hat{u}_i, \hat{\lambda}_i)$ are not consistent \citep{Jung2012,Lee2012a}.

Recall that $W_1 =  (\sigma_i z_{ij})_{i,j}$ is the $m \times n$ matrix of the scaled true scores, and $W_1 W_1\T $  is proportional to  the  $ m\times m$ sample covariance matrix of the first $m$   scores.  Similarly, we define
$\widehat{W}_1\T = d^{-1/2} \Xc(\hat{u}_1,\ldots,\hat{u}_m)$. Let $\{\varphi_i(S), v_i(S)\}$ denote the $i$th largest eigenvalue-eigenvector pair of a non-negative definite matrix $S$ and $v_{ij}(S)$ denote the $j$th loading of the vector $v_i(S)$.

\begin{theorem}\label{prop:samplescore}
Assume the $m$-factor model  under Conditions~\ref{A1}--\ref{A5} and let $n > m \ge 0$ be fixed. In addition, we assume the scores $w_{kj}$ are absolutely continuous.
\begin{itemize}
\item[(i)] If $k \le m$, then the ratio of the sample score to the true score of $X_j$ for the $k$th component is asymptotically decomposed into a common factor, not depending on $j$, and an error specific to each data point. Specifically,
     $$
     \frac{\hat{w}_{kj}}{w_{kj}} =  \rho_k  v_{kk}(W_1W_1\T ) + \varepsilon_{kj} + O_p(d^{-1/4})\quad  (j =1,\ldots,n),
     $$
where  $\rho_k = \{ 1+ \tau^2/\varphi_k(W_1W_1\T )\}^{1/2}$
and  $ \varepsilon_{kj} = \rho_k \sum_{1\le i \le m, i \neq k} {\sigma_i z_{ij}}({\sigma_k z_{kj}})^{-1}  v_{ki}(W_1W_1\T )$.
Moreover,
   \begin{align}\label{eq:SampleSCORErotation}
   \widehat{W}_1\T  = W_1\T R S  + O_p(d^{-1/4}),
   \end{align}
   where
   $R = [v_1(W_1W_1\T ),\ldots, v_m(W_1W_1\T )]$ is an $m\times m$ orthogonal matrix, and $S$ is the $m\times m$ diagonal matrix whose $k$th diagonal element is $ \rho_k $.

\item[(ii)] If $k > m $, then the ratio diverges with the rate $d^{(1- \gamma_k)/2}$, for $\gamma_k$ satisfying $\lambda_k \asymp d^{\gamma_k}$. Specifically,
       %\begin{align}\label{eq:divergingSCORERATE}
       ${\hat{w}_{kj}}/{w_{kj}} = O_p(d^{(1- \gamma_k)/2})$
       %\end{align}
       and $d^{-1}\sum_{j= 1}^n \hat{w}_{kj}^2 \to \tau^2$ in probability, as $d \to \infty$.
    \end{itemize}
\end{theorem}

The asymptotic relation (\ref{eq:SampleSCORErotation}) tells that for large $d$, the first $m$ sample scores in $\widehat{W}_1$ converges to the true scores in $W_1$, uniformly rotated and scaled for all data points.
It is thus valid to use the first $m$ sample principal scores for exploration of important data structure, to reduce the dimension of the data space from $d$ to $m$ in the high dimension, low sample size context.

\section*{Acknowledgement}
The authors are very grateful to the editor, associate editor and anonymous referees for their careful reading and thoughtful suggestions. % that help improve the manuscript greatly.

\section*{Supplementary material}
\label{SM}
Supplementary material available at \Bka\ online includes technical details and additional data examples.
%Further material such as technical details, extended proofs, code, or additional  simulations, figures and examples may appear online, and should be briefly mentioned as Supplementary Material where appropriate.  Please submit any such content as a PDF file along with your paper, entitled `Supplementary material for Title-of-paper\T .  After the acknowledgements, include a section `Supplementary material\T  in your paper, with the sentence , giving a brief indication of what is available.  Further instructions will be given when a paper is accepted.

\bibliographystyle{biometrika}

\bibliography{library}

%
%\newpage
%\includepdf[pages={1-25}]{NPCHDSupplement_20170816.pdf}

\end{document}